\documentclass[3p,times]{elsarticle}

\usepackage{amsmath}
\usepackage{amsthm}
\usepackage{booktabs}

\usepackage{amssymb}
\usepackage[figuresright]{rotating}
\usepackage{tcolorbox}

\usepackage[font=footnotesize,labelfont=bf]{caption}

\usepackage[english]{babel}
\usepackage{empheq,xcolor}

\begin{document}
	\begin{frontmatter}
		\title{Space Charge and Quantum Corrections in Free Electron Laser Evolution}

\author[Enea]{Giuseppe Dattoli}
\ead{giuseppe.dattoli@enea.it}

\author[Infn]{Hesham Fares}
\ead{hesham.fares@lnf.infn.it}

\author[Enea]{Silvia Licciardi\corref{cor}}
\ead{silvia.licciardi@enea.it}

\address[Enea]{ENEA - Frascati Research Center, Via Enrico Fermi 45, 00044, Frascati, Rome, Italy}
\cortext[cor]{Corresponding author: silviakant@gmail.com, silvia.licciardi@enea.it, orcid 0000-0003-4564-8866, tel. nr: +39 06 94005421. }
\address[Infn]{INFN, Via Enrico Fermi 40, 00044, Frascati, Rome, Italy and Assiut University, Department of Physics, 71515 Assiut, Arab Republic of Egypt.}

\begin{abstract}
Effects producing gain dilution in Free Electron Laser devices are well documented. We develop here a unified point of view allowing the introduction of space charge effects, along with the gain deterioration due to inhomogeneous broadening contributions and discuss the relevant interplay. We outline future developments and comment on the possibility of including in the formalism effects of quantum mechanical nature.
\end{abstract}


\end{frontmatter}

\section{Introduction}	
	
The space charge contributions to Free-Electron Laser ($FEL$) dynamics have been discussed in the past in a number of authoritative papers \cite{Colson,Sprangle,Sprangle2,Dermott,Gover}, which have clarified the physical mechanisms underlying the relevant bunching spoiling effects and the consequent gain degradation. \\

Shih and Yariv \cite{Shih} employed a single particle model and quantified the space ($SC$) charge detrimental effect in terms of the relativistic plasma frequency. The usefulness of their point of view stems from the fact that the afore mentioned quantity can be framed within the same context of the inhomogeneous broadening  parameters \cite{Dattoli,Dattoli7_2,Dattoli89,Dattoli7_3,Dattoli7_4,Xie,Xie_2,Xie_3,Russi,Russi_2}, used to specify the gain reduction induced by the electron beam relative energy spread and emittances. Even though the physical mechanisms underlying the two effects are not the same, the use of similar parameters allows a straightforward comparison of the relative detrimental consequences. We will comment on the relevant distinguishing physical features  in the final section of the paper. \\

\noindent In Refs. \cite{Gover,Shih} the authors treat the low gain case and write the gain function as
\begin{equation} \label{GrindEQ__1_} 
\begin{split}
& G\left(\phi ,\mu _{Q} \right)=2\pi g_{0} \frac{\phi }{\left( \phi ^{2} -\mu _{Q} {}^{2} \right) ^{2} } \left( 2-2 \cos (\phi ) \cos (\mu _{Q} )-\left( \frac{\mu _{Q} }{\phi } +\frac{\phi }{\mu _{Q} } \right) \sin (\mu_{Q} )\sin (\phi )\right), \\[1.1ex]
& \phi=2\pi N \;\dfrac{\omega_0 -\omega}{\omega_0},
\end{split} 
\end{equation} 
where $g_{0}$ is the small signal gain coefficient, $\phi $ is the detuning parameter and
\begin{equation} \label{GrindEQ__2_} 
\mu _{Q} =2\, N\, \left(\frac{\Omega _{p} \lambda _{u} }{2\, c} \right), \qquad \qquad \Omega _{p}^{2} =\frac{e^{2} \overline{N}_{e} }{\varepsilon _{0} m_{e} \gamma ^{3} } ,
\end{equation} 
with $N,\, \lambda _{u}$ being the number of undulator periods and period length respectively, $\overline{N}_{e}$ is the electron number density per unit volume, $m_{e}$ is the electron mass and $\gamma$ is the relativistic factor. A more appropriate way, for our purposes, of expressing $\mu _{Q}$ is the following
\begin{equation} \label{GrindEQ__3_} 
\mu _{Q} =\frac{2\, N}{\gamma } \, \left[\alpha \, \frac{E_{f} }{E_{e} } \frac{\lambda }{4\, \pi \, \varepsilon } \frac{\dot{N}_{e} \lambda _{u}^{2} }{c\beta _{T} } \right]^{\frac{1}{2} }
\end{equation} 
with
\begin{equation}\label{key}
\begin{split}
& \dot{N}_{e} =\frac{N_{e} }{\sqrt{2\, \pi } \sigma _{\tau } },\qquad \qquad  \qquad \qquad \quad 
E_{f} =\hbar \omega ,\\
& \lambda \equiv {\rm FEL\; wavelength}, \qquad \qquad \qquad \; E_{e} =m_{e} \gamma \, c^{2},\\
& \alpha \equiv {\rm fine\; structure\; constant}, \qquad \qquad  \beta _{{\rm T}} \equiv {\rm transverse\; twiss\; parameter}, \\
& \sigma _{\tau } \equiv {\rm electron\; bunch\; time\; duration} \;\;\;\;\; \varepsilon \equiv {\rm transvers \; beam\; emittance}.
\end{split}
\end{equation}
The physical content of $\mu _{Q}$ is clear, it states indeed that the $SC$ induced effects  decreases with increasing beam energy and increase for smaller $\sigma _{\tau }$ (namely for large peak currents) and smaller $\beta _{{\rm T}}$ decreasing transverse dimensions, hence larger density current.\\

\begin{figure}[h]
	\centering
	\includegraphics[width=0.5\linewidth]{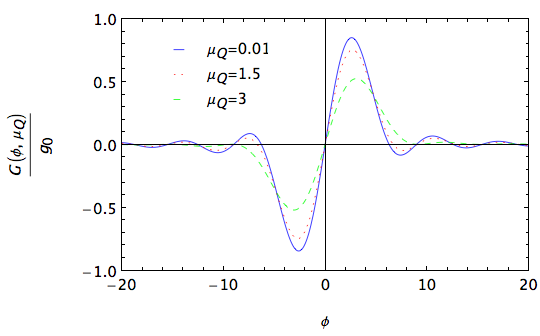}
	\caption{Gain divided $g_0$ vs. $\phi $ for different values of $\mu _{Q}$.}
	\label{fig:fig1}
\end{figure}

\noindent It is straightforwardly checked that, for vanishing $\mu _{Q} $, the gain function \eqref{GrindEQ__1_} reduces to the usual anti-symmetric form

\begin{equation} \label{GrindEQ__4_} 
\lim _{\mu _{Q} \to 0} \frac{G\left(\phi ,\mu _{Q} \right)}{g_{0} } =-\pi \, \partial _{\phi } \left[\dfrac{\sin \left(\dfrac{\phi }{2} \right)}{\left(\frac{\phi }{2} \right)} \right]^{2}  
\end{equation} 
The relevant consequences on the gain curve (a reduction of the peak and a broadening) are shown in Fig. \ref{fig:fig1}, where we have reported  $\frac{G\left(\phi ,\mu _{Q} \right)}{g_{0} }$ vs. $\phi$ for different values of $\mu _{Q}$.

\begin{figure}[h]
	\centering
	\includegraphics[width=0.4\linewidth]{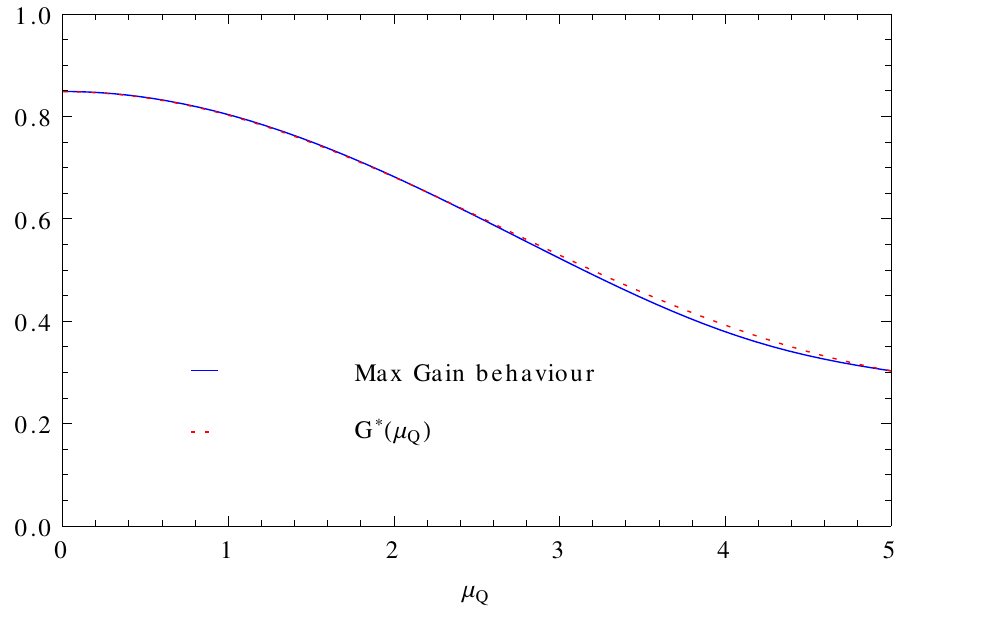}
	\caption{Comparison between maximum gain behaviour vs. $\mu _{Q}$ using numerical calculations for $\phi=0.01$ and scaling formula in Eq. \eqref{GrindEQ__5_}.}
	\label{fig:fig2}
\end{figure}

\noindent The presence of a non vanishing $\mu_Q$ determines a broadening and a shift of the gain line, which is not dissimilar from that due to the beam relative energy spread.  The behavior of the maximum gain vs. $\mu_Q$ is shown in Fig. \ref{fig:fig2} and is reproduced by the scaling relation
\begin{equation} \label{GrindEQ__5_} 
G^{*} \left(\mu _{Q} \right)\simeq 0.848g_{0} \exp \left( -\frac{\mu _{Q}^{2} }{18} +1.15\cdot 10^{-4} \mu _{Q}^{5} \right) .
\end{equation} 
It should be noted that Eq. \eqref{GrindEQ__5_} has not been derived on the basis of any physical assumption. The fitting formula has been chosen to get the best approximation with the numerical data. A Lorentzian type fitting formula, albeit less accurate, can also be exploited (See Section \ref{sec4} for more details).\\

\noindent We have closely followed the treatment of Shih and Yariv of Ref. \cite{Shih} and therefore we are restricted to the low gain case.\\

In the forthcoming sections we show how the study can be extended in various directions. 
\begin{enumerate}
	\item We will determine a fairly simple way of including in the gain formula the combined detrimental effects of space charge and relative energy spread. 
	\item We exploit the same formalism to go beyond the low gain approximation, used so far, and extend the treatment to the high gain regime, by a suitable modification of the relevant small signal integral equation.
	\item We eventually continue the discussion by proposing a modification to the equations ruling the $FEL$ small signal dynamics, which allows the inclusion of the quantum corrections \cite{DR,DRbis}
\end{enumerate}

\section{Low gain regime: Space charge and relative energy spread contribution}

The gain in Eq. \eqref{GrindEQ__1_} can be expressed in a different form more useful for our purposes such that the space charge effect can be displayed separately. We note indeed that it can be cast in the alternative way \cite{Shih}
\begin{equation} \label{GrindEQ__6_} 
G\left(\phi ,\mu _{Q} \right)=\dfrac{\pi \, g_{0} }{2\mu _{Q} } \left[\left(\dfrac{\sin \left(\dfrac{\phi _{-} }{2} \right)}{\dfrac{\phi _{-} }{2} } \right)^{2} -\left(\dfrac{\sin \left(\dfrac{\phi _{+} }{2} \right)}{\dfrac{\phi _{+} }{2} } \right)^{2} \right],\qquad \qquad \phi _{\mp } =\phi \mp \mu _{Q} .
\end{equation} 
According to the previous identity the gain function, including $\mu _{Q}$ corrections, has been expressed as the balance between two different emission processes. From the physical point of view it accounts for a kind of Raman effect including the frequency shift associated with the emission and absorption of a plasma oscillation (see also Ref. \cite{Freund} where this aspect of the problem is treated with high accuracy).\\

\begin{figure}[h]
	\centering
	\includegraphics[width=0.5\linewidth]{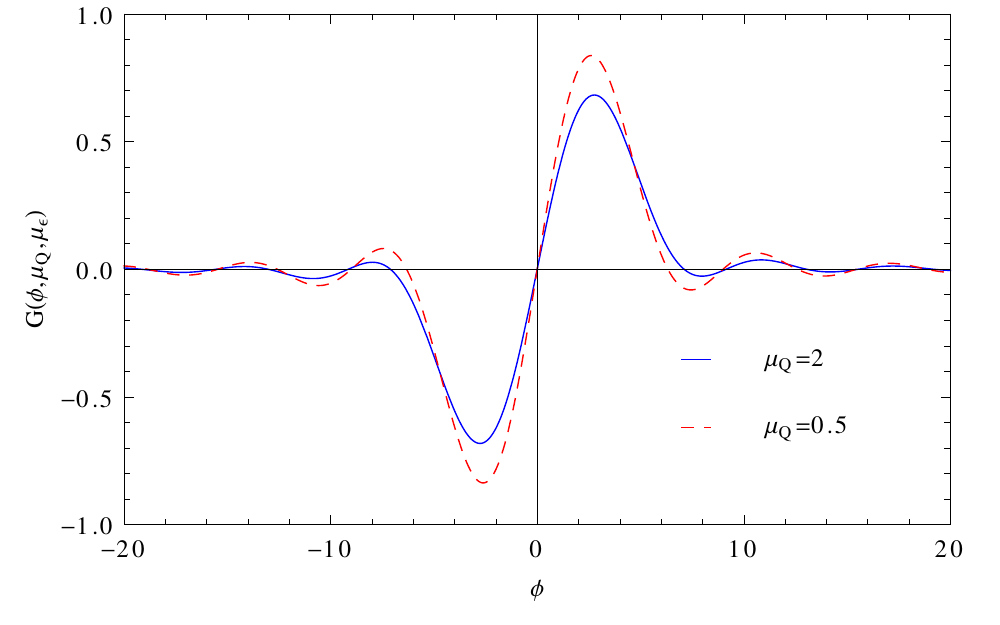}
	\caption{Gain function vs. $\phi$ with $\mu_{\varepsilon}=0.01$.}
	\label{fig:fig3a}
\end{figure}

\begin{figure}[h]
	\centering
	\includegraphics[width=0.5\linewidth]{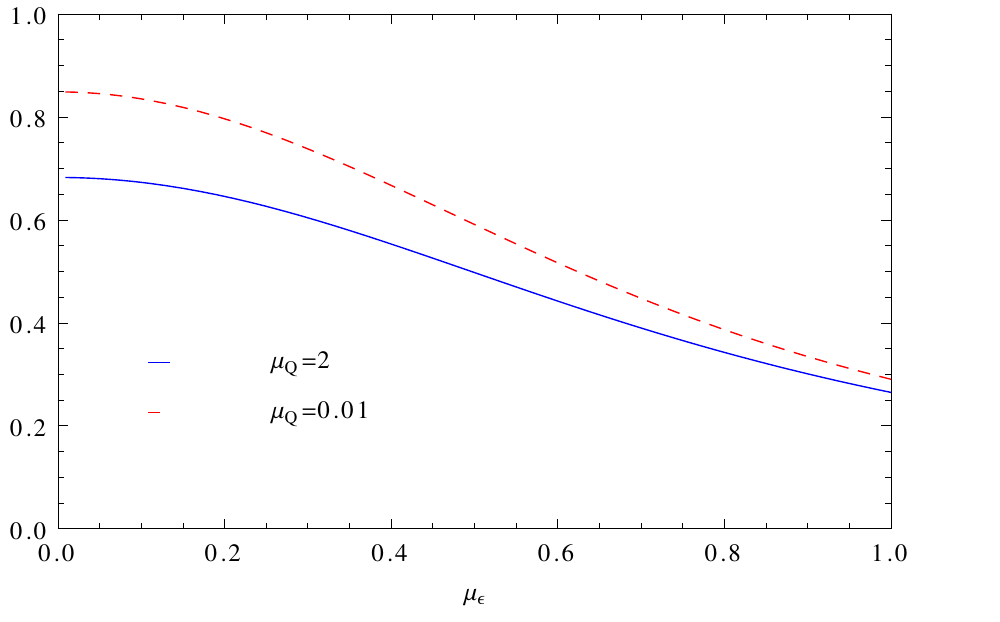}
	\caption{Maximum gain vs. $\mu_\varepsilon$ for different values of $\mu_Q$.}
	\label{fig:fig3b}
\end{figure}

\noindent The last equation contains the wave intensity variation only, if phase variations need to be included, we take advantage from the following integral representation of the spontaneous emission line 
\begin{equation} \label{GrindEQ__7_} 
S(\phi )=\left(\dfrac{\sin \left(\dfrac{\phi }{2} \right)}{\dfrac{\phi }{2} } \right)^{2} =2Re\left[E(\phi )\right], \qquad \qquad  \qquad
E(\phi )=\int _{0}^{1}(1-\xi )e^{-i\phi \, \xi }  d\xi .
\end{equation} 
The inclusion of the effects of the relative energy spread $\sigma _{\varepsilon }$ occurs through the introduction of the parameter \cite{Dattoli,Dattoli7_2,Dattoli89,Dattoli7_3,Dattoli7_4}
\begin{equation} \label{GrindEQ__8_} 
\mu _{\varepsilon } =4\, N\, \sigma_{\varepsilon }.  
\end{equation} 
After convolving the line-width \eqref{GrindEQ__7_} on a Gaussian energy distribution we get 
\begin{equation}\label{key}
E(\phi ,\mu _{\varepsilon } )=\int _{0}^{1}(1-\xi )e^{-i\phi \, \xi -\frac{\left(\pi\; \mu _{\varepsilon }\;\xi \right)^{2} }{2} }  d\xi 
\end{equation}
The combined effect of $\mu _{\varepsilon ,Q}$ parameters is eventually obtained as 
\begin{equation} \label{GrindEQ__10_} 
G\left(\phi ,\mu _{Q} ,\mu _{\varepsilon } \right)=\frac{\pi \, g_{0} }{\mu _{Q} } Re\left[E(\phi _{-} ,\mu _{\varepsilon } )-E(\phi _{+} ,\mu _{\varepsilon } )\right] .
\end{equation} 
In Fig. \ref{fig:fig3a} we have plotted the gain given by Eq. \eqref{GrindEQ__10_} vs. $\phi $ for different combination of $\mu _{Q,\varepsilon } $ and in Fig. \ref{fig:fig3b} the maximum gain vs. $\mu _{\varepsilon } $ for different values of $\mu _{Q} $. 

\noindent From Fig. \ref{fig:fig3a} it is worth noting that when the effect of energy spread increases (namely for larger values of $\mu_\varepsilon$), the space charge gain dilution becomes less effective. Accordingly we find that a good parameterization of the combined contributions is
\begin{equation} \label{GrindEQ__11_} 
G^{*} \left(\mu _{Q} ,\mu _{\varepsilon } \right)\simeq \dfrac{G^{*} \left(\dfrac{\mu _{Q} }{1+0.45\, \mu _{\varepsilon }^{2} } \right)}{1+1.7\, \mu _{\varepsilon }^{2} }  .
\end{equation} 
The previous scaling relation (reducing for vanishing $\mu _{Q}$ to the ordinary scaling vs. the energy spread \cite{Dattoli,Dattoli7_2,Dattoli89,Dattoli7_3,Dattoli7_4}, \cite{Renieri}) shows that $\mu _{\varepsilon }$ and $\mu_Q$ cannot be straightforwardly disentangled.\\

\noindent We have already noted that the results obtained so far are limited to the low gain regime. The inclusion of high gain effects will be accomplished in the forthcoming section using an appropriate modification of the high gain small signal integral equation.

\section{High Gain Regime and Space Charge Corrections}

The $FEL$ low gain condition is an approximation of the $FEL$ theory, which corresponds to the physical conditions in which the field amplitude can be kept constant, during the beam wave interaction inside the undulator. The complex amplitude and complex phase variations are registered at the end of the undulator. These conditions have characterized the early $FEL$ oscillator experiments, in which the small signal gain coefficient did not exceed few tens of percent. When high gain contributions become active, the field amplitude is ruled by a Volterra integro-differential equation with a memory kernel accounting for its self-consistent variation during the interaction.\\

\noindent Albeit the necessity for a more general formulation arouse very early \cite{Louisell}, the relevant analytical treatment was undertaken in the early eighties \cite{Schnitzer,Saldin,Yariv} and lead to the understanding of the role of the so called $FEL$ instability \cite{Bonifacio}, which, on the other side, is a characteristic feature of all the existing free electron generators of coherent radiation \cite{DGOR}.\\

\noindent In order to include the $SC$ corrections in the high gain formalism and write the corresponding Volterra equation, we consider the time dependent complex function 
\begin{equation} \label{GrindEQ__12_} 
E(\phi ,\tau )=-\phi\int _{0}^{\tau } \;e^{-i\phi \, \xi }  d\xi , \qquad \qquad \tau =\frac{z}{N\lambda _{u} } 
\end{equation} 
along with its extension
\begin{equation} \label{GrindEQ__13_} 
K\left( \phi ,\, \mu _{Q} ,\tau \right) =\frac{E(\phi _{-} ,\tau )-E(\phi _{+} ,\tau )}{2\, \mu _{Q} },  
\end{equation} 
including the space charge parameter.\\

\begin{figure}[h]
	\centering
	\includegraphics[width=0.5\linewidth]{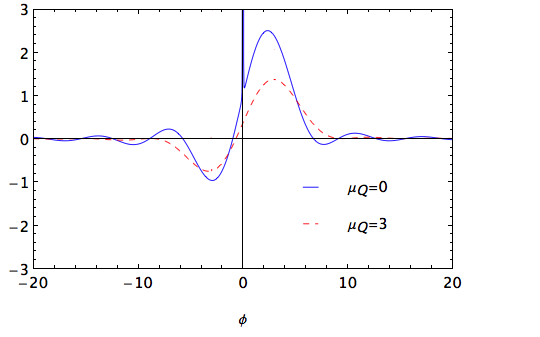}
	\caption{Gain vs. $\phi$ for $g_{0} =2$
		.}
	\label{fig:fig5}
\end{figure}

\noindent In correspondence of these quantities the high gain $FEL$ equation writes

\begin{equation} \label{GrindEQ__14_} 
\left\lbrace \begin{split}
& \frac{\partial }{\partial \tau } a=i\pi g_{0} \int _{0}^{\tau }K(\phi ,\mu _{Q} ,\tau ')\, a(\tau -\tau ')d\tau '  ,\\
& a(0)=1
\end{split}\right. 
\end{equation} 
where $a(\tau )$ is the $FEL$ dimensionless Colson amplitude.Before proceeding further let us note that Eq. \eqref{GrindEQ__14_} can also be written as
\begin{equation} \label{GrindEQ__15_} 
\frac{\partial }{\partial \tau } a=i\pi g_{0} \frac{e^{-\mu _{Q} \frac{\partial }{\partial \phi }} -e^{\mu _{Q} \frac{\partial }{\partial \phi } } }{2\mu _{Q} } \int _{0}^{\tau}E(\phi ,\tau ')\, a(\tau -\tau ')d\tau '  
\end{equation} 
for vanishing $\mu _{Q}$ we find 
\begin{equation}\label{key}
\lim _{\mu _{Q} \to 0} \frac{e^{-\mu _{Q} \frac{\partial }{\partial \phi } } -e^{\mu _{Q} \frac{\partial }{\partial \phi } } }{2\mu _{Q} } =-\frac{\partial }{\partial \phi }
\end{equation}
and under this assumption Eq. \eqref{GrindEQ__15_} reduces to the ``canonical'' $FEL$ high gain equation \cite{Dattoli81} 
\begin{equation}\label{id}
\frac{\partial }{\partial \tau } a=i\pi g_{0} \int _{0}^{\tau }e^{-i\phi \tau '} \, \tau 'a(\tau -\tau ')d\tau ' .
\end{equation}
In Fig. \ref{fig:fig5} we have shown the gain function for values of the small signal gain parameter inducing high gain corrections ($g_{0} =2$) for cases with and without space charge effects. We have checked the correctness of the numerical procedure and we have found that the maximum gain exhibits the scaling vs. $g_{0}$ provided by the identity \cite{Centioli}
\begin{equation} \label{GrindEQ__17_} 
G^{*} (g_{0} )\simeq 0.848\, g_{0} +0.19\cdot g_{0}^{2} +4.23\cdot 10^{-3} g_{0}^{3}  
\end{equation} 
The variables in Eq. \eqref{id} are more suitable for the low gain case. A more convenient form for the high regime is provided by
\begin{equation}\label{zLg}
\partial_{\tilde{z}}\;a=\dfrac{i}{3\sqrt{3}}\int_{0}^{\tilde{z}}e^{-i\tilde{\phi}\tilde{z}'}\tilde{z}'a(\tilde{z}-\tilde{z}')d\tilde{z}' , \qquad \qquad
\tilde{z}=\dfrac{z}{L_g}
\end{equation}
where
\begin{equation}\label{key}
\tilde{\phi}=\dfrac{1}{2\rho\sqrt{3}}\dfrac{\omega_0-\omega}{\omega_0}, \qquad \qquad 	L_{g} =\frac{\lambda _{u} }{4\, \pi \, \sqrt{3} \rho }  
\end{equation}
with $\rho$ being the Pierce parameter linked to the small signal gain coefficient by\footnote{It should be noted that, being $g_0 \propto N^3$, there is no dependence in Eq. \eqref{zLg} on the number of periods of the undulator.}
\begin{equation}\label{key}
\rho=\dfrac{\left( \pi g_0\right) ^{\frac{1}{3}}}{4\pi N}
\end{equation}
and $L_g$ being the gain length which specifies the growth rate of the high gain $FEL$'s operating in single pass configuration. 

\noindent  Also, in the limit of high gain, the presence of space charge contributions yields a gain distortion not dissimilar from those induced by the energy spread inhomogeneous broadening.\\

Non ideal beam qualities determines an increase of $L_{g}$ and of the saturation length as well. Within this respect relative energy spread becomes harmful if the condition $\sigma _{\varepsilon } \le \frac{\rho }{2} $ is not fulfilled. In the high gain regime the parameter controlling the inhomogeneous effect is $\tilde{\mu }_{\varepsilon } =\frac{2\, \sigma _{\varepsilon } }{\rho }$ and the corresponding quantity for the $SC$ is
\begin{equation}\label{tmuq}
\tilde{\mu }_{Q} =\frac{1}{\gamma \, \rho } \, \left[\alpha \, \frac{E_{f} }{E_{e} } \frac{\lambda }{4\, \pi \, \varepsilon } \frac{\dot{N}_{e} \lambda _{u}^{2} }{c\beta _{T} } \right]^{\frac{1}{2} }.
\end{equation}
An analogous quantity\footnote{In Ref. \cite{Marcus} the $SC$ parameter is defined by replacing $\gamma^3$ in the second of Eqs. \eqref{GrindEQ__2_} with $\gamma\gamma_z^2$, where $\gamma_z=\dfrac{\gamma}{\sqrt{1+\frac{K^2}{2}}}$.} has been introduced in Ref. \cite{Marcus}, it has been normalized using the same criterion for the quantities marking the inhomogeneous broadening effects and quoting the authors of \cite{Marcus} ``\textit{the $SC$ parameter is scaled to be twice the plasma phase advance over a one-dimensional gain length}".\\
We can now comment on the possibility that these effects be observed in an actual experimental configuration, or, saying better, whether they may induce negative sizeable effects like the increase of the saturation length in e.g. $FEL$ designed  as sources of bright hard $X$-ray beams. They demand for high charges, short bunches and high quality beams. A paradigmatic example is provided by  the European $X\!-\!FEL$, which is foreseen to produce photons up to the angstrom wavelength, with a beam of a $17.5$ $GeV$ produced by a linear accelerator \cite{Zagorodnov}. The high current intensity is achieved by compressing bunches bearing a charge of $250$ and $500$ $pC$. To this aim the electron beam line incorporates three vertical bunch compressors of $C$ type \cite{Zagorodnov}.\\

\noindent It is evident that the major problem in handling this type of beams is the control the stability of the compression and the qualities of the electron bunch after the compression.\\

\noindent The analysis of these problems goes beyond the scope of this paper, we assume therefore that compressing a $500$ $pc$ to hundreds of fs does not create additional problems in terms of energy spread and emittance (for more substantive comments see Ref. \cite{Zagorodnov}).\\

\begin{figure}[h]
	\centering
	\includegraphics[width=0.5\linewidth]{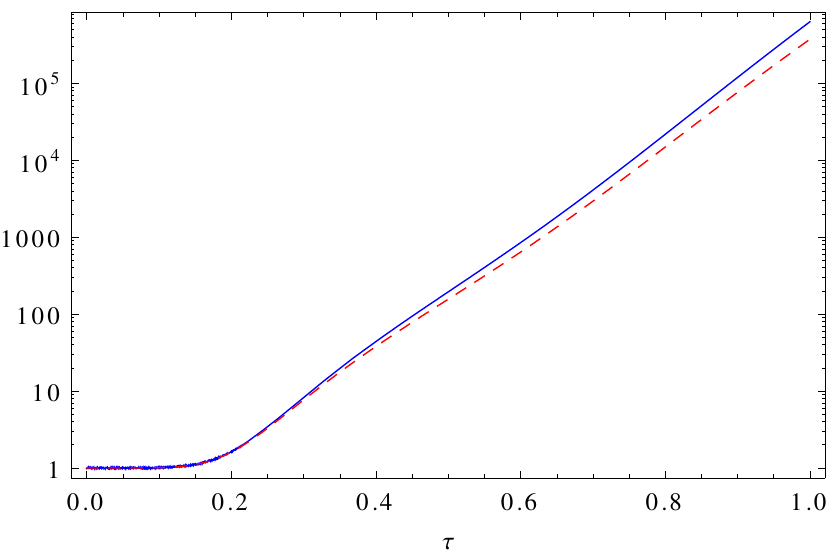}
	\caption{Dimensionless intensity growth $\mid a \mid^2$ vs. $\tau$ for $\mu_Q=0.24$ (blue) and $\mu_Q=3$ (red dashed) with $g_0=200$.}
	\label{fig:fig6m}
\end{figure}

\noindent Before entering further into the discussion whether the space charge effects may have some relevance in high $FEL$ devices by keeping in consideration the previously quoted $X\!-\!FEL$ parameters, we note that $\tilde{\mu }_{Q} \simeq 3$ may slightly modify the $FEL$ intensity growth along the undulator, as reported in Fig. \ref{fig:fig6m}. 

\noindent If we assume that the produced $X$-ray-beam is diffraction limited (namely $\varepsilon \simeq \frac{\lambda }{4\, \pi } $) we can write
\begin{equation} \label{GrindEQ__20_} 
\tilde{\mu }_{Q} =\frac{5\cdot 10^{-6} }{\gamma \, \rho } \, \left[\, \frac{E_{f} }{E_{e} } \frac{\dot{N}_{e} \lambda _{u}^{2} }{\beta _{T} } \right]^{\frac{1}{2} }  .
\end{equation} 
Imposing the condition  $\tilde{\mu }_{Q} >3$ to appreciate some space charge distortions to the $FEL$ dynamics, we obtain the following condition on the peak beam current
\begin{equation} \label{GrindEQ__21_} 
\hat{I}=e\, \dot{N}_{e} >5.76\cdot 10^{11} \left(\gamma \rho \right)^{2} \frac{\beta _{T} }{\lambda _{u}^{2} } \frac{E_{e}. }{E_{f} }  
\end{equation} 
Keeping for example $E_{e} \simeq 15$ $GeV$, $E_{f} \simeq12.4$ $keV$, $\lambda _{u} =0.04\;m$, $\beta _{T} =8\;m$ and assuming $\gamma \rho \simeq 5$ we get $\hat{I}>5\;kA$. According to Ref. \cite{Zagorodnov} operating conditions with $10\;kA$ are foreseen, by considering such an extreme value we can assume values of $\tilde{\mu }_{Q} $ on the order of $5$ which (as reported in Fig. \ref{fig:fig5}) might allow quite a significant increase of the gain length induced by space charge effects. In the following section we will provide further comments on $SC$ impact on $FEL$ device and on their interplay with other effects including those of quantum nature.

\section{Final considerations}\label{sec4}

In the previous parts of the article, we have discussed a fairly simple method to embed space charge and inhomogeneous broadening effects into a straightforward procedure. The method we have adopted to extend the analysis to the high gain seems to rely on a heuristic recipe, because we just exploited the function $E(\phi ,\tau )$ to derive the high gain integral equation. Albeit we have checked the consistency of the procedure with previous results, in absence of a more rigorous mathematical tool there are still elements which makes our analysis doubtful.\\

\noindent We have therefore checked our results by making a comparison with two previous papers in which the problem is addressed reducing the high gain equation to a third order differential equation, which in our notation reads
\begin{equation} \label{GrindEQ__22_} 
a'''+2\, i\, \tilde{\phi} \, a''+(\tilde{\mu} _{Q}^{2} -\tilde{\phi} ^{2} )a'=\dfrac{i}{3\sqrt{3}}\; a 
\end{equation} 
where the apices denote derivatives taken with respect to $\tilde{z}$.\\

The use of the procedure envisaged in Ref. \cite{DiPalma} allows to transform Eq. \eqref{GrindEQ__22_}  into an integro-differential equation. We note indeed that we can factorize the differential operators on the l.h.s. as 
\begin{equation}\label{key}
\left(\partial_{\tilde{z}} + i \phi_+ \right) \left(\partial_{\tilde{z}} + i \phi_- \right) \left(\partial_{\tilde{z}} a\right)=\dfrac{i}{3\sqrt{3}}\;a 
\end{equation}
which can also be written as
\begin{equation}\label{key}
\partial_{\tilde{z}} a=\dfrac{i}{3\sqrt{3}}\hat{K}a, \qquad \qquad \hat{K}=\left[  \left(\partial_{\tilde{z}} + i \phi_+ \right)\left(\partial_{\tilde{z}} + i \phi_- \right)\right] ^{-1}.
\end{equation}
The operator $\hat{K}$ is easily shown to be an integral operator, we note indeed that
\begin{equation}\label{key}
\left[  \left(\partial_{\tilde{z}} + i \phi_+ \right) \left(\partial_{\tilde{z}} + i \phi_- \right)\right] ^{-1}=-\dfrac{i}{2\tilde{\mu }}_Q\left(\dfrac{1}{\partial_{\tilde{z}} + i \phi_+}-\dfrac{1}{\partial_{\tilde{z}} + i \phi_-} \right) 
\end{equation}
The use of standard techniques of Laplace transform eventually reduces its action on the dimensionless amplitude $a$ to the integral form in Eq. \eqref{GrindEQ__14_} .\\

\begin{figure}[h]
	\centering
	\includegraphics[width=0.5\linewidth]{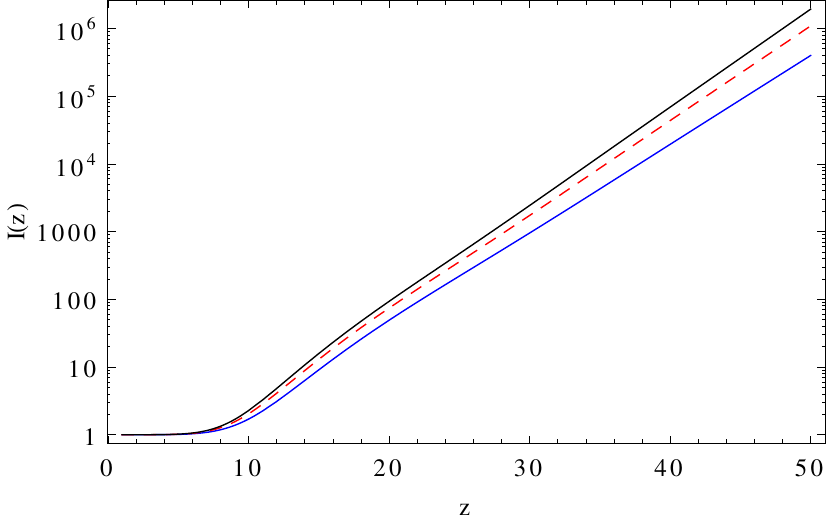}
	\caption{Dimensionless intensity growth vs. $z$ for different values of $\tilde{\mu }_{Q} $ ($0$ black, $3$ dashed red, $5$ blue).}
	\label{fig:fig7}
\end{figure}

We have already mentioned that the presence of the $SC$ contributions produces changes similar to those of inhomogeneous broadening and indeed the most significant effect which can be drawn from Fig. \ref{fig:fig6m} is an increase of the gain length, which can be cast in the form
\begin{equation}\label{eq29}
L_g(\tilde{\mu}_Q)=\dfrac{\lambda_u}{4\pi\sqrt{3}\rho(\tilde{\mu}_Q)}, \qquad \qquad \qquad
\rho(\tilde{\mu}_Q)=\dfrac{\rho}{1+0.974\left(\dfrac{\tilde{\mu}_Q}{16} \right)^2+0.983 \left(\dfrac{\tilde{\mu}_Q}{16} \right)^4}
\end{equation}
In the high regimethe use of a rational function appears more appropriate to get an accurate fit (a maximum relative error within $2\%$) of $SC$ effect on the Pierce parameter \cite{Dattoli7_3}. An analogous fitting formula has been proposed in Ref. \cite{Marcus} using a power law scaling, in the spirit of Ming Xie parameterization \cite{Xie,Xie_2,Xie_3}.The use of the analytical form for the linear intensity growth, along with the redefinition of \eqref{eq29} of the gain length
\begin{equation}\label{contr}
I(z)=\dfrac{I_0}{9}\left[3+2\cosh\left(\dfrac{z}{L_g\left( \tilde{\mu}_Q\right)} \right)+4\cos\left(\dfrac{\sqrt{3}}{2}\dfrac{z}{L_g\left( \tilde{\mu}_Q\right)} \right) \cosh\left(\dfrac{z}{2L_g\left( \tilde{\mu}_Q\right)} \right)  \right] 
\end{equation}
yields a fairly accurate reproduction of the numerical results reported in Fig. \ref{fig:fig7}.\\

\noindent We can however gain further insight by solving Eq. \eqref{GrindEQ__22_} with, e.g., $\phi=0$. Assuming that the solution be of the exponential type $a\propto e^{i\lambda\tilde{z}}$, the exponents $\lambda$ are obtained as the roots of the cubic equation 
\begin{equation}\label{key}
\lambda^3-\tilde{\mu}_Q^2\lambda+\dfrac{1}{3\sqrt{3}} =0. 
\end{equation}
The field driven by the fast growing root behaves like  
\begin{equation}\label{key}
\begin{split}
& a\propto e^{\Lambda\tilde{z}},\\
& 
\Lambda=\dfrac{\sqrt{3}}{2}\sqrt[3]{-\dfrac{1}{6\sqrt{3}}+\sqrt{\dfrac{1}{\left(6\sqrt{3} \right)^2 }-\dfrac{\tilde{\mu}_Q^6}{27}}}-
\sqrt[3]{-\dfrac{1}{6\sqrt{3}}-\sqrt{\dfrac{1}{\left(6\sqrt{3} \right)^2}-\dfrac{\tilde{\mu}_Q}{27}^6}}, \qquad \;\; \tilde{\mu}_Q<\left(  \dfrac{1}{4}\right)^{\frac{1}{6}}.
\end{split}
\end{equation}
The gain length, affected by space charge effects, can therefore be defined as 
\begin{equation}\label{key}
L_g\left(\tilde{\mu}_Q \right)=\dfrac{L_g}{\Lambda} 
\end{equation}
which is consistent (albeit in a limited interval of $\tilde{\mu}_Q$) with what is expected from Eq. \eqref{eq29}.\\
The inclusion of $SC$ effects on the $FEL$ dynamics has been based on the exam of the induced distortion of the gain function, within the context of a $1$D analysis. The inclusion of diffractive effects can be accomplished by substituting the $\rho$ parameter with its $3$D counterpart, according to the Ming-Xie recipe or to that in \cite{Dattoli7_4}. The equations ruling the evolution of the beam transverse section (with respect to the longitudinal coordinate) is affected by a transverse Coulomb force and reads (for a round beam)
\begin{equation}\label{key}
\sigma^{{}_{''}}=\dfrac{\varepsilon}{\sigma^3}+\dfrac{1}{4}\dfrac{\Omega_p^2}{c^2}\dfrac{\sigma_0^2}{\sigma}
\end{equation}
with $\sigma_0$ being the $SC$ unperturbed beam section (absence of space charge contributions). The increase of the beam transverse dimension $\sigma$ and divergence $\sigma'$ contribute to the $FEL$ gain detriment through a decrease of the current density and  with a further inhomogeneous broadening effect associated with the $SC$ induced emittance increase. The pivotal parameter allowing the quantifications of these further effects is $\tilde{\mu}_Q$ as shown in a forthcoming investigation.\\
The final point we like to touch is whether it is possible to include in this scheme the so called quantum corrections \cite{DR,DRbis,DR2}, which have raised recently interest, mainly for the works in \cite{Debus,Moritz}.
Albeit we will examine the interplay between classical and quantum parameter affecting $FEL$ dynamics more thoroughly in a forthcoming note, we just discuss whether the procedure put forward so far is suitable to include this further contribution.\\

To this aim we remind that the relevance of quantum effect is measured by the parameter

\begin{equation}\label{muq}
\tilde{\mu}_q=\dfrac{\hbar \omega}{\rho m_e \gamma c^2}
\end{equation}
which is recognized as quantity determining the ratio between the classical energy spread $\Delta \gamma m_e c^2$ and the energy of an emitted photon $\hbar \omega$ where the induced energy spread $\Delta\gamma \sim\rho\gamma$. It is obvious that when $\tilde{\mu}_Q<1$ the discreteness of energy exchange becomes significant and the quantum effect becomes apparent, where a distinct transition between two energy levels is possible. For properly detuning of the system, each electron emits one photon at maximum. A physical counterpart of the Self-Amplified Spontaneous Emission ($SASE$) $FEL$ operated in a quantum regime is the photons scattering from electrons in the Compton backscattering \cite{etal}. In a first approximation, considering only the number of photon growth along the undulator, we find that it is ruled by an equation of the type \eqref{GrindEQ__22_} with $\tilde{\mu}_q$ in place of $\tilde{\mu}_Q$ \cite{DR,DRbis,etal,Schroeder}.\\ 

\noindent The gain process can accordingly be viewed as the balance between two processes regarding the emission and absorption of a photon of energy $\hbar\omega$.\\

The associated dispersion relation can therefore be written as
\begin{equation}\label{key}
a^{{}_{'''}}+2i\tilde{\phi}a^{{}_{''}}+\left(\tilde{\mu}_Q^2+\tilde{\mu}_q^2-\tilde{\phi}^2 \right)a^{{}_{'}}=\dfrac{i}{3\sqrt{3}}\;a. 
\end{equation}
Quantum and $SC$ terms affect the $FEL$ intensity evolution in the same way,  which indicates that it is quite hard to disentangle the effects if one is just looking at the field intensity evolution. 
The possibility of observing effects of quantum nature is a very difficult task. Ad hoc designed experimental configurations, like those reported in 
\cite{Debus,etal}, should be accurately examined in order to disentangle the quantum contributions from all the others (homogeneous broadenings and $SC$) contributing to the $FEL$ gain. \\
%
%


This point, even though needing substantive improvements, indicates that the formalism we have followed is suitable to address the inclusion of effects of different nature in $FEL$ dynamics, within the same unifying context. The procedure, even though hampered by its $1$-D nature, may provide a first and useful aid to get a feeling of their importance and whether they can be experimentally detected.\\

\textbf{Acknowledgements}\\

The present work is part of the CompactLight project and has received funding from the European
Union's Horizon 2020 research and innovation programme in the framework of the project
under grant agreement No. 777431.\\

\textbf{Author Contributions}\\

Conceptualization: G.D., H.F.; methodology: G.D., H.F.; data curation: G.D., S.L.; validation: G.D., H.F., S.L.; formal analysis: G.D., H.F., S.L.; writing - original draft preparation: G.D., H.F..; writing - review and editing: S.L. .\\


\textbf{References}


\begin{thebibliography}{}
	
	\bibitem{Colson} W.B. Colson, The nonlinear wave equation for higher harmonics in free-electron lasers, \textit{IEEE J. Quantum Electr.}, vol. 17, no. 8,  1981.
	
	\bibitem{Sprangle} P. Sprangle, C.M. Tang, W.M. Manheimer, Nonlinear Formulation and Efficiency Enhancement of Free-Electron Lasers, \textit{Phys. Rev. Lett.}, 43, published 1979. 
	
	\bibitem{Sprangle2} P. Sprangle, C.M. Tang, W.M. Manheimer, Nonlinear theory of free-electron lasers and efficiency enhancement, \textit{Phys. Rev. A}, 21, 302, 1980.
	
	\bibitem{Dermott} B. Mc Dermott, T.C. Marshall, The collective free-electTon laser, \textit{Phys. Quantum Electr.}, vol. 7, Addison-Wesley, pp. 509--522, 1980.
	
	\bibitem{Gover} A. Gover, Z. Livni, Operation regimes of Cerenkov-Smith-Purcell free electron lasers and T. W. amplifiers, \textit{Opt. Commun.}, vol. 26, pp. 375--380, 1978. 
	
	\bibitem{Shih} C. C. Shih and A. Yariv, Single-electron analysis of the space-charge effect in free-electron lasers, \textit{Phys. Rev. A}, vol. 22, pp. 2217--2222, 1980.
	
	\bibitem{Dattoli} G. Dattoli, T. Letardi, J.M.J. Madey, A. Renieri, Limits on the Single-Pass Higher Harmonics FEL Operation, \textit{J. Quantum Electron.}, vol. 20, 9, pp. 1003--1005, 1984. 
	
	\bibitem{Dattoli7_2} G. Dattoli, A. Renieri, A.Torre, R. Caloi, Inhomogeneous broadening effects in high-gain free electron laser operation: A simple parametrization, \textit{Il Nuovo Cimento D}, 11, pp. 393--404, 1989. 
	
	\bibitem{Dattoli89} G. Dattoli, H. Fang, L. Giannessi, M. Richetta, A. Torre, R. Caloi, Parametrizing the gain dependences in a single passage FEL operating with moderate current e-beams, \textit{Nuclear Instruments and Methods in Physics Research Section A: Accelerators, Spectrometers, Detectors and Associated Equipment}, vol. 285, 1--2, pp. 108--114, 1989.
	
	\bibitem{Dattoli7_3} G. Dattoli, P.L. Ottaviani, S. Pagnutti, Booklet for FEL design:aA collection of practical formulae, Frascati: ENEA–Edizioni Scientifiche, 2007, http://www.fel.enea.it/booklet-presentation.html .
	
	\bibitem{Dattoli7_4} G. Dattoli, L. Giannessi, P.L. Ottaviani, C. Ronsivalle, Semi-analytical model of self-amplified spontaneous-emission free-electron lasers, including diffraction and pulse-propagation effects, \textit{J. Appl. Phys.}, vol. 95, pp. 3206--3210, 2004.
	
	\bibitem{Xie} M. Xie, Exact and variational solutions of 3D eigenmodes in high gain FEL's, \textit{Nuclear Instruments and Methods in Physics Research Section A: Accelerators, Spectrometers, Detectors and Associated Equipment}, vol. 445, 1--3, pp. 59--66, 2000.  
	
	\bibitem{Xie_2} M. Xie, Design Optimization for an X-Ray Free Electron Laser Driven by SLAC Linac, Proceedings Particle Accelerator Conference 1995, http://accelconf.web.cern.ch/AccelConf/p95/ARTICLES/TPG/TPG10.PDF .
	
	\bibitem{Xie_3} K.J. Kim , M. Xie, Self-amplified spontaneous emission for short wavelength coherent radiation, \textit{Nuclear Instruments and Methods in Physics Research Section A: Accelerators, Spectrometers, Detectors and Associated Equipment A}, 331, 1--3, pp. 359--364, 1993. 
	
	\bibitem{Russi} E.L. Saldin, E.A. Schneidmiller, M.V. Yurkov, Design Formulas for VUV and X-Ray FELs, MOPOS15 in Proceedings of the 26th Free Electron Laser Conference, pp. 139--142,Trieste, Italy, 2004.
	
	\bibitem{Russi_2} E.L. Saldin, E.A. Schneidmiller, M.V. Yurkov, \textit{The Physics of Free Electron Lasers}, Springer, 2000, ISBN 978-3-662-04066-9.
	
	\bibitem{DR} G. Dattoli, A. Renieri, Chapter: Quantum theory of free electron laser in \textit{Laser Handbook}, Volume 6: Free electron Laser by W.B. Colson et al., North Holland, 1990.
	
	\bibitem{DRbis} G. Dattoli, A. Renieri, Chapter: Experimental and theoretical aspects of the free-electron laser in \textit{Laser Handbook}, Volume 4: Free electron Laser by M.L. Stitch and M. Bass, North Holland, 1990.
	
	\bibitem{Freund} H.P. Freund, T.M. Jr Antonsen,  Principles Of Free Electron Lasers, 3rd ed., Springer, 2018, ISBN 10: 3319751050.
	
	\bibitem{Renieri} G. Dattoli, A. Renieri, A. Torre, Lectures on the Free Electron Laser Theory and Related Topics, World Scientific, 1993.
	
	\bibitem{Louisell} W.H. Louisell, J.F. Lam, D.A. Copeland, W.B. Colson, Exact classical electron dynamic approach for a  free-electron laser amplifier, \textit{Phys. Rev. A}, vol. 19, pp. 288--300, 1979. 
	
	\bibitem{Schnitzer} I. Schnitzer, A. Govers, The prebunched free electron laser in various operating gain regimes, \textit{Nuclear Instruments and Methods in Physics Research Section A: Accelerators, Spectrometers, Detectors and Associated Equipment},  237, pp. 124--140, 1985.
	
	\bibitem{Saldin} E.L. Saldin, E.A. Schneidmiller, M.V. Yurkov, The features of an FEL oscillator with a tapered undulator, \textit{Opt. Commun.}, 103, 1993. 
	
	\bibitem{Yariv} C.C. Shih, A. Yariv, Inclusion of space  charge  effects  with  Maxwell’s  equations  in  the  single  particle  analysis of free electron lasers, \textit{IEEE J. Quantum  Electron.}, vol. QE-17, pp. 1387--1394,  1981. 
	
	\bibitem{Bonifacio} R. Bonifacio, C. Pellegrini, L.M. Narducci, Collective instabilities and high-gain regime in a free electron laser, \textit{Optics Commun.}, vol. 50, 6, pp. 373--378, 1984. 
	
	\bibitem{DGOR} G. Dattoli, E. Di Palma, S. Pagnutti, E. Sabia, Free Electron Coherent Sources: From Microwaves to X-rays, \textit{Phys Rep.}, vol. 739, pp. 1--51, 2018.
	
	\bibitem{Dattoli81} G. Dattoli, A. Marino, A. Renieri, F. Romanelli, Progress in the Hamiltonian picture of the free-electron laser, \textit{IEEE J.f Quantum Electr.} vol. 17, 8,  1981. 
	
	\bibitem{Centioli} G. Dattoli, A. Torre, C. Centioli, M. Richetta, Free Electron Laser, Operation in the intermediate gain region, \textit{IEEE J-QE}, 25, 2327, 1989.
	
	\bibitem{Marcus} G. Marcus, E. Hemsing, J. Rosenzweig, Gain length fitting formula for free-electron lasers with strong space-charge effects, \textit{Phys. Rev. Special Topics - Accelerators and Beams} 14, 080702, 2011.
	
	\bibitem{Zagorodnov} I. Zagorodnov, M. Dohlus, S. Tomin, Accelerator beam dynamics at the European X-ray Free Electron Laser, \textit{Phys. Rev. Special Topics- Accelerators and Beams}, 22, 024401, 2019.
	
	\bibitem{DiPalma} G. Dattoli, E. Di Palma, F. Nguyen, E. Sabia, Generalized Trigonometric Functions and Elementary Applications,  \textit{Int. J. Appl. Comput. Math}, 3, pp. 445--458, 2017. 
	
	\bibitem{DR2} G. Dattoli, F. Nguyen, Free Electron Laser and Fundamental Physics, \textit{Progress in Particle and Nuclear Physics}, Elsevier, vol. 99, pp. 1--28, 2018.
	
	\bibitem{Debus}  A. Debus, K. Steiniger, P. Kling, C. Moritz Carmesin, R. Sauerbrey, Realizing quantum free-electron lasers: a critical analysis of experimental challenges and theoretical limits, \textit{Phys. Scr.}, vol. 94, 7, 2019.
	
	\bibitem{Moritz} C. Moritz Carmesin, P. Kling, E. Giese, R. Sauerbrey, W.P. Schleich,
	Quantum and classical phase-space dynamics of a free-electron laser, arXiv 1911.12584v1 [quant-ph] 28 Nov 2019.
	
	\bibitem{etal} R. Bonifacio, H. Fares, M. Ferrario, B. W.J. McNeil, G.R.M. Robb, Design of sub-Angstrom compact free-electron laser source, \textit{Opt.Commun.}, 382, pp. 58--63, 2017. 
	
	\bibitem{Schroeder} C.B. Schroeder, C. Pellegrini, P. Chen, Quantum effects in high-gain free-electron lasers, \textit{Phys. Rev. E}, 64, 056502, 2001.
	
	
\end{thebibliography}
\end{document}